\RequirePackage{lineno}

\documentclass[floatfix,superscriptaddress,a4paper,
               showpieces,showkeys,nofootinbib,preprint]{revtex4-2}

\usepackage{epsfig}
\usepackage{latexsym}
\usepackage{xspace}
\usepackage{comment}
\usepackage{soul}
\usepackage{inputenc}
\usepackage{indentfirst}
\usepackage{enumerate}
\usepackage{color}
\usepackage[dvipsnames]{xcolor}
\usepackage{epstopdf}
\usepackage{colortbl}
\usepackage{blindtext}
\usepackage[normalem]{ulem}

\usepackage{amsmath}
\usepackage{bm}
\usepackage[english]{babel}
\usepackage{url}
\usepackage{hyperref}

\graphicspath{{plots/}}

\begin{document}
%%%%%%% units
\newcommand{\kg}{\ensuremath{\mbox{kg}}\xspace}
\newcommand{\eV}{\ensuremath{\mbox{e\kern-0.1em V}}\xspace}
\newcommand{\GeV}{\ensuremath{\mbox{Ge\kern-0.1em V}}\xspace}
\newcommand{\MeV}{\ensuremath{\mbox{Me\kern-0.1em V}}\xspace}
\newcommand{\GeVc}{\ensuremath{\mbox{Ge\kern-0.1em V}\!/\!c}\xspace}
\newcommand{\GeVcc}{\ensuremath{\mbox{Ge\kern-0.1em V}\!/\!c^2}\xspace}
\newcommand{\AGeV}{\ensuremath{A\,\mbox{Ge\kern-0.1em V}}\xspace}
\newcommand{\AGeVc}{\ensuremath{A\,\mbox{Ge\kern-0.1em V}\!/\!c}\xspace}
\newcommand{\MeVc}{\ensuremath{\mbox{Me\kern-0.1em V}/c}\xspace}
\newcommand{\T}{\ensuremath{\mbox{T}}\xspace}
\newcommand{\cmsq}{\ensuremath{\mbox{cm}^2}\xspace}
\newcommand{\msq}{\ensuremath{\mbox{m}^2}\xspace}
\newcommand{\cm}{\ensuremath{\mbox{cm}}\xspace}
\newcommand{\mm}{\ensuremath{\mbox{mm}}\xspace}
\newcommand{\micron}{\ensuremath{\mu\mbox{m}}\xspace}
\newcommand{\mrad}{\ensuremath{\mbox{mrad}}\xspace}
\newcommand{\ns}{\ensuremath{\mbox{ns}}\xspace}
\newcommand{\m}{\ensuremath{\mbox{m}}\xspace}
\newcommand{\s}{\ensuremath{\mbox{s}}\xspace}
\newcommand{\ms}{\ensuremath{\mbox{ms}}\xspace}
\newcommand{\ps}{\ensuremath{\mbox{ps}}\xspace}
\newcommand{\dd}{\ensuremath{{\textrm d}}\xspace}
\newcommand{\dedx}{\ensuremath{\dd E\!/\!\dd x}\xspace}
\newcommand{\tof}{\ensuremath{\textup{\emph{tof}}}\xspace}
\newcommand{\pt}{\ensuremath{p_{\textrm T}}\xspace}
\newcommand{\PT}{\ensuremath{P_\textup{T}}\xspace}
\newcommand{\mt}{\ensuremath{m_{\textrm T}}\xspace}

%particles
\newcommand{\pbar}{\ensuremath{\overline{\textup{p}}}\xspace}
\newcommand{\p}{\ensuremath{\textup{p}}\xspace}
\newcommand{\nbar}{\ensuremath{\overline{\textup{n}}}}
\newcommand{\dbar}{\ensuremath{\overline{\textup{d}}}}
\newcommand{\pim}{\ensuremath{\pi^-}\xspace}
\newcommand{\pip}{\ensuremath{\pi^+}\xspace}
\newcommand{\km}{\ensuremath{\textup{K}^-}\xspace}
\newcommand{\kp}{\ensuremath{\textup{K}^+}\xspace}

% inverse hyperbolic functions
%\DeclareMathOperator{\acosh}{acosh}
%\DeclareMathOperator{\asinh}{asinh}
%\DeclareMathOperator{\atanh}{atanh}

%%%%%%%%%%%%% some software programs and generators
%----- software
\def\Offline{\mbox{$\overline{\text%
{Off}}$\hspace{.05em}\raisebox{.4ex}{\underline{line}}}\xspace}
\def\SHOE{\mbox{SHO\hspace{-1.34ex}\raisebox{0.2ex}{\color{green}\textasteriskcentered}\hspace{0.25ex}E}\xspace}
\def\DSHACK{\mbox{DS\hspace{0.15ex}$\hbar$ACK}\xspace}
\DeclareRobustCommand{\SHINE}{\mbox{\textsc{S\hspace{.05em}\raisebox{.4ex}{\underline{hine}}}}\xspace} %DeclareRobustCommand allows this to work in caption
%\def\SHINE{\textsc{Shine}\xspace}

%----- event generators
\def\Glissando{\textsc{Glissando}\xspace}
\newcommand{\FlukaLong}{{\scshape Fluka2008}\xspace}
\newcommand{\FlukaEleven}{{\scshape Fluka2011}\xspace}
\newcommand{\Fluka}{{\scshape Fluka}\xspace}
\newcommand{\UrqmdLong}{{\scshape U}r{\scshape qmd1.3.1}\xspace}
\newcommand{\Urqmd}{{\scshape U}r{\scshape qmd}\xspace}
\newcommand{\GheishaLong}{{\scshape Gheisha2002}\xspace}
\newcommand{\GheishaOld}{{\scshape Gheisha600}\xspace}
\newcommand{\Gheisha}{{\scshape Gheisha}\xspace}
\newcommand{\Corsika}{{\scshape Corsika}\xspace}
\newcommand{\Venus}{{\scshape Venus}\xspace}
\newcommand{\VenusLong}{{\scshape Venus4.12}\xspace}
\newcommand{\GiBUU}{{\scshape GiBUU}\xspace}
\newcommand{\GiBUULong}{{\scshape GiBUU1.6}\xspace}
\newcommand{\FlukaNewLong}{{\scshape Fluka2011.2\_17}\xspace}
\newcommand{\Root}{{\scshape Root}\xspace}
\newcommand{\Geant}{{\scshape Geant}\xspace}
\newcommand{\GeantThree}{{\scshape Geant3}\xspace}
\newcommand{\GeantFour}{{\scshape Geant4}\xspace}
\newcommand{\QGSJet}{{\scshape QGSJet}\xspace}
\newcommand{\DPMJet}{{\scshape DPMJet}\xspace}
\newcommand{\Epos}{{\scshape Epos}\xspace}
\newcommand{\EposLong}{{\scshape Epos1.99}\xspace}
\newcommand{\QGSJetLong}{{\scshape QGSJetII-04}\xspace}
\newcommand{\DPMJetLong}{{\scshape DPMJet3.06}\xspace}
\newcommand{\SibyllLong}{{\scshape Sibyll2.1}\xspace}
\newcommand{\EposLHCLong}{{\scshape EposLHC}\xspace}
\newcommand{\Hsd}{{\scshape Hsd}\xspace}
\newcommand{\Ampt}{{\scshape Ampt}\xspace}

\newcommand{\NASixtyOne}{NA61\slash SHINE\xspace}

\newcommand{\U}{{\mathcal{U}}\xspace}
\newcommand{\V}{{\mathcal{V}}\xspace}
\newcommand{\C}{{\mathcal{C}}\xspace}

%%%%%%%%%%%%%%%%%%%%%%%% misc
\def\red#1{{\color{red}#1}}
\def\blue#1{{\color{blue}#1}}
\def\magenta#1{{\color{magenta}#1}}
\def\green#1{{\color{green}#1}}

\def\avg#1{\langle{#1}\rangle}
\def\sci#1#2{#1\!\times\!10^{#2}}
\newcommand{\Fi}[1]{Fig.~\ref{#1}}

\newcommand{\const}{\mathop{\rm const}\nolimits}
\newcommand{\bs}{\boldsymbol}
\newcommand*{\dis}{\displaystyle}
\newcommand{\bvar}{\mathbf}
\newcommand{\eq}[1]{\begin{align} #1 \end{align}}

\title{
Spatial correlations of charm and anticharm quarks at hadronisation
}

\author{M. Gazdzicki}
%\affiliation{Goethe-University Frankfurt am Main, Germany}
\affiliation{Jan Kochanowski University, Kielce, Poland}
\author{D. Kiko\l{}a}
\affiliation{Warsaw University of Technology, Warsaw, Poland}
\author{I. Pidhurskyi}
%\affiliation{Goethe-University Frankfurt am Main, Germany}
\affiliation{Jan Kochanowski University, Kielce, Poland}
\affiliation{European Organization for Nuclear Research, CERN, Geneva, Switzerland}
\author{L. Tinti}
\affiliation{Jan Kochanowski University, Kielce, Poland}

\begin{abstract}
% --------------------------
%         Abstract
% --------------------------
Heavy-ion collisions are a unique tool for studying properties of strong interactions at high energy densities. In particular, the momentum correlations of charm and bottom hadrons have been considered for testing heavy quark thermalisation in the dense matter produced by the collisions. 

In this respect, two effects have been considered: the decrease of the initial back-to-back correlations and the increase of correlations due to heavy-quark interactions with the collectively flowing medium.
 
Here, we show that information on the spatial correlations of the charm-anticharm quarks at the hadronisation can be extracted by measuring the momentum correlation of charm and anticharm hadrons produced in central collisions of two heavy nuclei. This, however, requires collisions with a single charm-anticharm quark pair created - the condition likely to be fulfilled in central Pb+Pb collisions at the CERN SPS energies.
We introduce a method to correct the measured joint distribution function for the smearing of the charm and anticharm hadron momenta caused by hadronisation. 
Then the results are directly sensitive to the spatial correlations at the hadronisation.
Using an example of central Pb+Pb collisions at the CERN SPS energies, we demonstrate that even a limited statistics of charm-anticharm hadron pairs can distinguish between different spatial correlation functions of charm-anticharm quarks at hadronisation.

The results on spatial charm-anticharm quark correlations will provide a unique 
test of different assumptions on heavy quark creation in space-time and transport in dense, strongly interacting matter. We show that the existing detector technology and beam intensities at the CERN SPS should allow us to conduct the needed experiments soon.

\end{abstract}

\maketitle

\newpage
%\linenumbers

% -----------------------
%        Text 
% -----------------------

\section{Introduction}

Collisions of heavy ions at relativistic energies provide insights into fascinating features of nuclear matter at high energy densities.
This includes the creation and properties of the Quark-Gluon Plasma (QGP)~\cite{Shuryak:1980tp} - a state of matter with quark and gluon degrees of freedom expected to exist in the Universe's 
first moments. Moreover, there is a possibility of discovering the critical point of strongly interacting matter; for example, see Refs.~\cite{Gazdzicki:2015ska, Bzdak:2019pkr} and references therein.
Impressive progress has been made in experimental and theoretical studies in the last decades. Still, many properties of matter at high densities and particle-antiparticle creation in the medium remain to be uncovered. 

Measurements of correlations between a charm meson and its antiparticle have been proposed to test the equilibration of charm~\cite{Zhu:2006er,Cao:2015cba} in momentum space. In a semi-classical picture, the initial back-to-back momentum correlations between the $c$ and $\bar c$ quarks are reduced by the interactions with the medium and
hadronisation of the quarks; see, for example, Ref.~\cite{He_2020} and references therein.
Thus, the charm hadron correlations provide means for quantifying transport properties of the strongly interacting medium, complementary to measurements of collective 
effects~\cite{Song:2007fn} and modification of momentum spectra via nuclear modification factor $R_{\mathrm{AA}}$~\cite{Kharzeev:2004yx}.

In this paper, we present and discuss different physics that can be addressed by studying momentum correlations between charm and anticharm hadrons. 
Using azimuthal correlations of charm and anticharm hadrons, one can study the spatial correlations of charm and anticharm quarks
at hadronisation. Specifically, we show that the observed correlations provide direct insights into whether heavy quarks hadronise close to each other in the coordinate space or if hadronisation points are distant. 
This information shall allow testing different assumptions on the creation mechanism of heavy quarks and antiquarks and their transport in the quark-gluon plasma. The creation mechanism is the key input assumption in charm and bottom quark interaction models with the quark-gluon plasma. All the modern experiments at SPS, RHIC, and the LHC are conducted with programs that aim to quantify the QGP parameters using heavy quarks.

The new idea presented in this paper  
utilises the collective flow of charm hadrons measured
in heavy-ion collisions at high energies~\cite{STAR:D0:v1,STAR:D0:v2,STAR:eHF:v2:2014yia,ALICE:D0:v2,ALICE:D0:v2:2013,ALICE:eHF:v2,ALICE:D:v2:2020iug,ATLAS:muHF:v2:2018ofq,ALICE:D:v2:2017pbx,ALICE:beauty:v2:2020hdw,PHENIX:eHF:v2}. For clarity of the presentation we assume that final-state momenta of charm and anticharm hadrons are given by the superposition of the charm quark flow and a contribution due to quarks' hadronisation. Other possible effects, like back-to-back correlation of created heavy quarks, influencing charm hadron momenta, are also discussed.
\iffalse
Hadronic rescattering 
and final-state interactions are neglected, supported by the recent measurements of interaction parameters of $D$-mesons with hadrons~\cite{ALICE:2022enj}.
\fi
The flow contribution depends uniquely on the hadronisation point. Thus, one can extract the charm quark hadronisation point by measuring the quark flow and having the space-time dependence of the flow at hadronisation. The latter information can be obtained by adjusting the flow models to hadron-production results measured in the same reaction. The same concerns the anticharm hadronisation point. Consequently, one gets the spatial distance between $c$ and
$\bar{c}$ quarks at the hadronisation by measuring the difference between their flows at hadronisation.

Experiments measure momenta of charm and anticharm hadrons instead of the wanted flow components. The paper introduces an unfolding method 
to overcome this problem for central collisions obeying isotropy in the transverse plane. 

Here, we stress that the above strategy is valid independently of how many $c$ and $\bar{c}$ quarks are created in a single collision. 
We, however, argue that it is important to study the simplest case, in which only one charm and anticharm quark pair is created in a collision. This will allow explicitly testing the commonly accepted postulate of the $c-\bar{c}$ creation close in space-time and their transport with subluminal velocities.
Interpreting results on collisions with many $c$ and $\bar{c}$ quarks would require model input on multi-quark correlations. 
On average, one expects about three $c-\bar{c}$ pairs in central Au+Au collisions at $\mathrm{\sqrt{s_{NN}} = 200}$~GeV (RHIC)~\cite{STAR:2018zdy,STAR:2012nbd}, and a few tens $c-\bar{c}$ pairs in central Pb+Pb reactions at $\mathrm{\sqrt{s_{NN}} = 5.02}$ (LHC)~TeV~\cite{ALICE:2021dhb,ALICE:2021rxa}.
The condition of only one $c$ and $\bar{c}$ pair in a collision is approximately fulfilled at the CERN SPS energies~\cite{Merzlaya:2024cbt}.
For this reason, we consider an example of central Pb+Pb collisions at the CERN SPS energies. 
This example can be straightforwardly extended to bottom hadron production at RHIC or the LHC.

The heavy-quark production and azimuthal correlations in heavy-ion collisions at very high energies were addressed theoretically in the past; for a review, see Ref.~\cite{Andronic:2015wma}. In particular, they were considered as a tool for uncovering a mechanism behind the jet suppression~\cite{Attems:2022ubu,Attems:2022otp} and the study of 
charm energy-loss mechanism~\cite{Rohrmoser:2017vsa,PhysRevC.90.024907,Wang:2019vhg,Wang:2021xpv}.
The heavy-quark spatial diffusion in QCD matter was discussed recently in 
Refs.~\cite{Sambataro:2020pge,Capellino:2022nvf,Satapathy:2022xdw}, see also references therein.
The ATLAS experiment measured the azimuthal-angle correlations of muon pairs originating from heavy-flavour decays in  Pb+Pb collisions at 5.02~TeV~\cite{ATLAS:2022mhn}. One notes that the measured 
muon pairs come from jet-like correlations of heavy-flavour hadrons at high transverse momenta.
Our work addresses, for the first time, the possibility of studying the spatial correlations of heavy quarks and antiquarks at hadronisation.  

The paper is organised as follows.
First, we briefly discuss theoretical challenges in predicting ab initio hadron correlations in
heavy-ion collisions, Sec.~\ref{sec:challenges}.
Then, the qualitative idea of extracting the spatial correlation of charm and anticharm quarks at hadronisation is
quantified using simple modelling presented in Sec.~\ref{sec:model}. The procedure to unfold smearing due to hadronisation is also introduced in this section. The section closes by giving arguments for the importance of measurements of collisions with only one charm and anticharm quark created. The feasibility of the corresponding experiments is discussed in Sec.~\ref{sec:experiment}, and the results are summarised in Sec.~\ref{sec:summary}. Additional information is included in the Appendix.

\section{Theoretical challenges}
\label{sec:challenges}

Quantum Chromodynamics (QCD) is the commonly accepted theory of strong interactions.
However, attempts to derive precise quantitative predictions for
multi-particle production in high-energy collisions from QCD have been unsuccessful. Predictions of QCD–inspired models suffer from
uncertainties that are difficult to estimate. Here, we discuss them in aspects relevant to this work.

The most popular QCD-inspired approaches to predict hadron production in heavy-ion collisions are based on classical approximations. For instance, this is the case for the relativistic kinetic theory and hydrodynamic models~\cite{Florkowski:Phenomenology}. 
In heavy-ion collisions, the quantum effects are expected to be large or even comparable to the classical predictions, at least regarding the flow of energy and momentum~\cite{Tinti:2023mtv}. This is because the typical action scale of the system, a few hundred (at most)~MeV of temperature and spatial changes by a fraction of a femtometer, is smaller than $\hbar c \simeq 200$ MeV$\cdot$fm. The surprising success of hydrodynamical models in describing nuclear reactions can probably be traced back either to 
%
%still be understood regardless of the validity of the relativistic kinetic theory and %for large gradients and deviations from equilibrium,
%
the attractor dynamics~\cite{Jankowski:2023fdz} or the generalized off-shell hydrodynamic expansion~\cite{Tinti:2023kpq}. In any case, these arguments hold only for the hydrodynamic variables and not, for instance, for the two-particle correlations.
Operators' expectation values, such as the energy density in hydrodynamics, are generally considered. Still, their fluctuations (e.g., variance and higher-order moments) and related correlations are more difficult to deal with. 

It is important to stress that, despite the success in predicting some observables, the approximations used in quantitative models are inadequate for the full data description. For instance, hydrodynamics cannot properly address spectra at high transverse momenta. Relativistic kinetic theory assumes molecular chaos and removes two and three-particle correlations already at the classical level. Most physically proven quantum effects (diffraction, entanglement, etc.) are neglected and cannot be addressed by the current models. It is not simple to estimate the size of the quantum effects, lacking quantitative models that include them.
Thus, whether the correlations produced by a classical treatment like relativistic kinematics are enough to describe the experimental results is unclear. 
For further discussion, see Appendix~A.

Considering the above,
we test whether extreme assumptions on $c$ and $\bar{c}$ spatial and momentum correlations lead to experimentally distinguishable predictions.
If yes, the experimental results should distinguish between models based on these assumptions.

%It is not considered in publicly available event generators and models of heavy quark interaction with partonic matter like AMPT\cite{Wang:2019vhg,Wang:2021xpv} or JETSCAPE framework~\cite{JETSCAPE:2022hcb} either. 

%---------------------------------------------------------------------
\section{Quantitative predictions and discussion}
\label{sec:model}

The following assumptions are made to quantify the intuitive expectations concerning the relation of the spatial and momentum correlations of $c$, $\bar{c}$ quarks and $D$, $\bar{D}$ hadrons at hadronisation:

\begin{enumerate}[(i)]
\item
The production of charm and anticharm hadrons in central Pb+Pb collisions is considered.
The collision energy is assumed to be adjusted to have a mean charm multiplicity below one, allowing for neglecting production of more than one $c$- and $\bar{c}$-hadron pair in a single collision. This likely corresponds to the top CERN SPS energy ($\sqrt{s_{NN}} \approx 17$~GeV)~\cite{Snoch:2018nnj, Merzlaya:2024cbt}.

\item 
The charm and anticharm hadrons are emitted from the hadronisation hypersurface of a spherical fireball, which is parametrised as:
\begin{equation}
\label{eq:hypersurface}
    t^2 = r^2 + \tau_{\mathrm{HAD}}^2~,
\end{equation}
where $r = \sqrt{x^2 + y^2 + z^2}$ is the distance from the centre of the fireball ($\mathbf{r} = ({x,y,z})$) and $\tau_{\mathrm{HAD}}$ is the hadronisation proper time (natural units are used). The distance is assumed to be limited to $r\le R_{\mathrm{MAX}}$.
For simplicity,
the parameters $\tau_{\mathrm{HAD}}$ and $R_{\mathrm{MAX}}$
are assumed to be equal to the Pb nucleus radius: $\tau_{\mathrm{HAD}} = R_{\mathrm{MAX}}= 6$~fm. The sketch shown in Fig.~\ref{fig:hypersurface} illustrates the assumptions.

\begin{figure}[ht]

\includegraphics[width=0.7\textwidth]{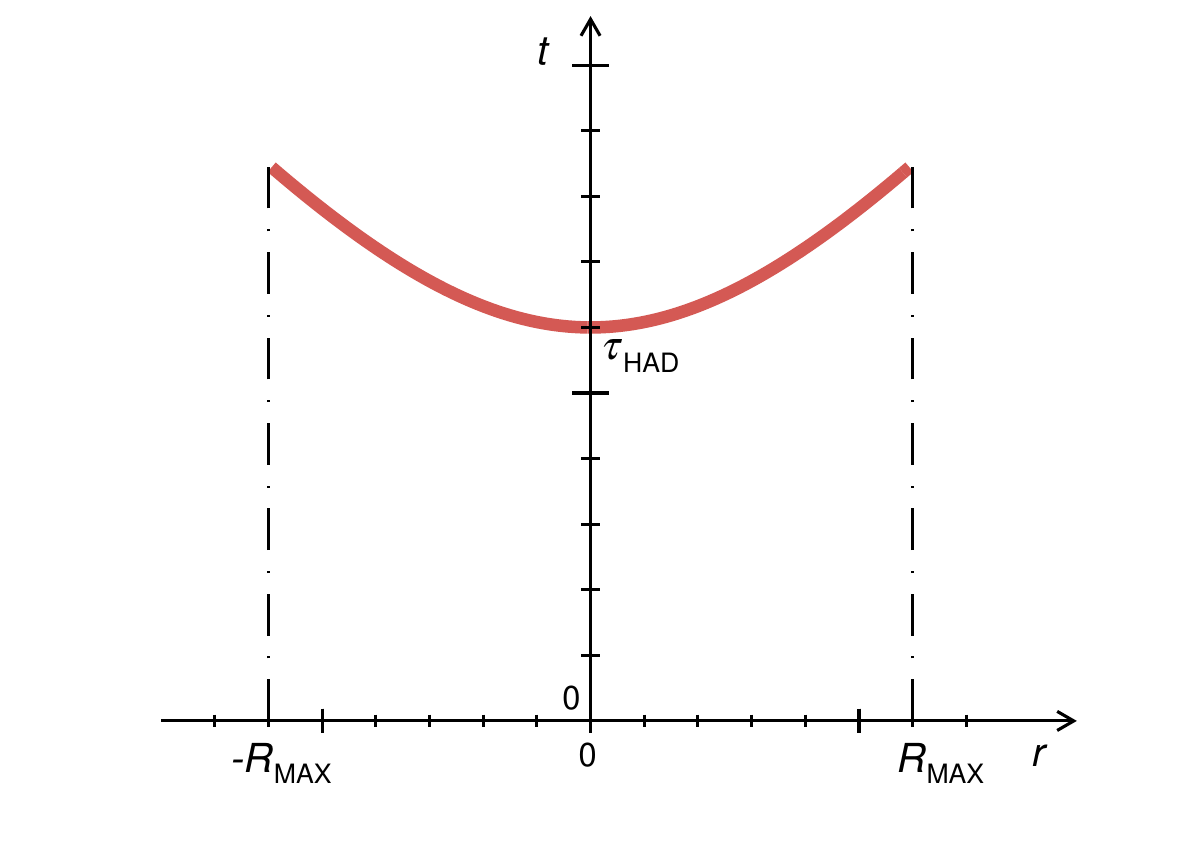}

\caption{
The sketch illustrating the assumed hadronisation hypersurface and its parameters, see text and Eq.~\ref{eq:hypersurface} for details.
}
\label{fig:hypersurface}
\end{figure}

\item
The four-velocity of the flowing matter at the hadronisation is assumed to be 
\begin{equation}
\label{eq:flow}
u^\mu =x^\mu/\tau_{\mathrm{HAD}}~,
\end{equation}
where $x^{\mu}$ is the hadronisation space-time point. This Hubble-like proportionality of the velocity to the distance is frequently used in modelling heavy-ion collisions~\cite{Tsegelnik:2022eoz}.

\item For simplicity, it is assumed that the emission probability of (anti)charm hadrons averaged over collisions is uniform on the hadronisation hypersurface. This is consistent with the assumption of the spherical fireball and the constant hadronisation temperature (see the next point).

\item
In the rest frame of the flow, the (anti)charm hadron momentum $\mathbf{p}$ distribution at the hadronisation hypersurface is assumed to be the statistical one:
\begin{equation}
\label{eq:had}
   \frac{\mathrm{d}^3N} {\mathrm{d}p~\mathrm{d}^2\Omega} ~\propto~ p^2 \cdot \exp{ \left( - \frac{ \sqrt{m^2 + p^2} }{T_{\mathrm{HAD}}} \right) }~,
\end{equation}
where $m = 1.869$~\GeVc is the charm hadron mass assumed to be equal to the $D^0$ meson mass, and the temperature parameter 
is $T_{\mathrm{HAD}}$~=~150~MeV~\cite{Andronic:2017pug}. 
The hadronisation momenta of charm hadrons are drawn independently.
\item
The obtained hadronisation four-momentum is boosted with the flow velocity to calculate the hadron momentum in the collision rest frame.
\end{enumerate}

Then, the results on momentum correlations between $c$ and $\bar{c}$ quarks (hadrons) are calculated assuming three $c - \bar{c}$ spatial correlation functions at hadronisation:
\begin{enumerate}[(a)]
\item
The  $c$ and $\bar{c}$ quarks hadronisation points are identical.
The spatial correlation function is $\delta$-like.
Thus, their flow components are identical.
Still, the corresponding $c-$ and $\bar{c}-$hadron momenta differ because of 
the independent hadronisation. 
\item
The  $c$ and $\bar{c}$ quarks hadronisation points are uncorrelated.
The spatial correlation function is uniform.
Thus, their flow components are also uncorrelated. 
The corresponding $c-$ and $\bar{c}-$hadron momenta are also uncorrelated as the hadronisation components are independent. 
\item
The intermediate case is modelled by drawing the pair hadronisation point according to the uniform distribution on the hadronisation hypersurface and then drawing independently $c$ and $\bar{c}$ hadronisation points according to the 3D Gauss distribution centred at the pair hadronisation point and having $\sigma = \sigma_x = \sigma_y = \sigma_z =2$~fm. Then, the hadronisation times of $c$ and $\bar{c}$ are calculated to ensure the points are at the hypersurface.
The spatial correlation function is Gaussian-like.
The flow components of $c$- and $\bar{c}$ quarks are different but correlated, leading to the correlation of charm and anticharm hadron momenta. 
\end{enumerate}

In general, the two-particle distribution function depends on six momentum components of 
\( \mathbf{p}_1 \) and \( \mathbf{p}_2 \) momentum vectors, where
the subscripts $_{1}$ and $_{2}$ stand for charm and anticharm hadrons, respectively.
The symmetries reduce the number of non-trivial arguments of the distribution function. In the model, we consider only the case of central heavy-ion collisions, which are isotropic in the transverse plane ($x-y$)~\cite{NA49:2003njx} - invariant under rotation in the 
azimuthal angle.  
The experimental results of these collisions indicate that the system at hadronisation is elongated along the direction of the beam ($z$). Considering this, we discuss the model predictions only in the transverse plane. 

In the transverse plane, the two-particle distribution depends on four components of two transverse momentum vectors.
Due to the azimuthal symmetry of the model, this dependence reduces to three non-trivial momentum quantities. Here, we select them as:
\begin{itemize}
\renewcommand{\labelitemi}{-}
    \item The opening angle between the transverse momentum vectors, 
    \begin{equation}
    \label{eq:Phi}
    \mathit{\Phi} = min(~|\phi_1 - \phi_2|,~  2 \pi - |\phi_1 - \phi_2|~)~,
    \end{equation}    
    where \( \phi_1 \) and \( \phi_2 \) are the azimuthal angles of charm and anticharm particles (quarks or hadrons) changing in the range 
    \( [-\pi, \pi] \). By definition $\mathit{\Phi}$ changes between 0 and \( \pi \).
    \item The transverse momentum vectors' magnitudes \( p_{\textrm{T}, 1} \), and
    \( p_{\textrm{T}, 2} \).    
\end{itemize}

\begin{figure}[ht]
\includegraphics[width=0.90\textwidth]{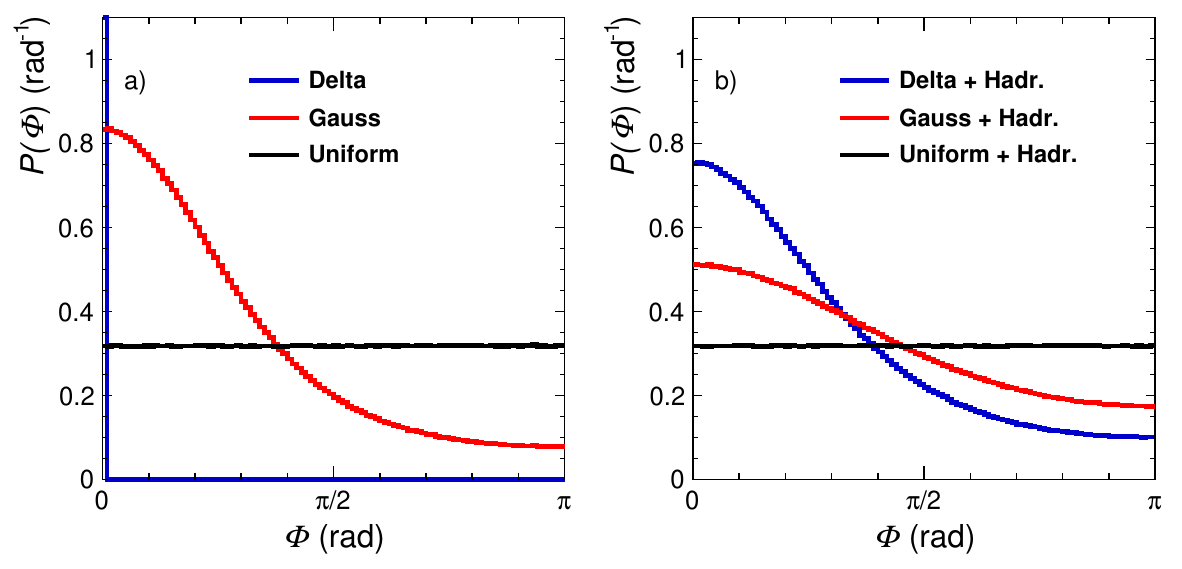} 
\caption{
Distribution of $c-\bar{c}$ quark (the left plot) and hadron (the right plot) pairs at hadronisation 
in the azimuthal opening angle, $\mathit{\Phi}$ 
calculated using three spatial $c-\bar{c}$ correlation function: $\delta$-like (Delta), 3D Gauss with $\sigma_x = \sigma_y = \sigma_z =2$~fm (Gauss) and uniform (Uniform). The left plot shows predictions for the correlation function due to the quarks' flow. The predictions shown on the right plot are calculated for momenta of $D$ and 
$\bar{D}$ hadrons, including the flow and the hadronisation components.
}
\label{fig:DPhi}
\end{figure}

Figure~\ref{fig:DPhi} shows the distribution of $c-\bar{c}$ quark (the left plot) and hadron (the right plot) pairs at hadronisation 
in the azimuthal opening angle, $\it{\Phi}$. The distribution was obtained by integrating over all values of \( p_{\textrm{T}, 1} \), 
and \( p_{\textrm{T}, 2} \). The results are obtained using the Monte Carlo technique with $10^7$ events generated. The left plot shows predictions for the correlation function due to the quarks' flow. Due to the assumed azimuthal angle symmetry, the opening angle obtained using the flow velocities is identical to the opening angle calculated using the hadronisation points, see Appendix~B.
Thus, the correlation function in the spatial azimuthal opening angle equals the correlation function in $\it{\Phi}$, the flow opening angle.
Note that this is not the case for peripheral heavy-ion collisions. Due to the anisotropy in the transverse plane, the flow opening angle generally significantly differs from the spatial opening angle, see Appendix~B.

Results shown on the right plot of Fig.~\ref{fig:DPhi} are obtained for momenta of $D$ and $\bar{D}$ hadrons, including the flow and the hadronisation components.
One observes that the effect of the hadronisation on
the  $\mathit{\Phi}$ significantly depends on the assumed spatial correlation between $c$ and $\bar{c}$ quarks at hadronisation.  
The uniform distribution in $\it{\Phi}$ remains uniform after the hadronisation independently of the flow and hadronisation modelling.
The hadronisation significantly smears the $\delta$-like spatial correlation of the quark hadronisation points.
The broader the spatial quark correlation, the smaller the change due to the hadronisation.  

We note that the flow correlation function in $\mathit{\Phi}$ has a maximum at zero and is uniform in the limit of uncorrelated hadronisation points. It is qualitatively different from the back-to-back momentum correlations expected for the $c-\bar{c}$ creation in hard processes at the early stages of collisions~\mbox{\cite{Zhu:2006er,Cao:2015cba}} leading to a maximum at $\mathit{\Phi} = \pi$. Thus, the correlation function in $\mathit{\Phi}$ corrected for the hadronisation may have a saddle shape with the maxima at zero and $\pi$, if both sources of correlations are present. This should allow us to take into account the initial back-to-back correlations when extracting the spatial correlation of $c$ and $\bar{c}$ quarks at hadronisation. 
We expect that experimental data will allow us to distinguish between these two causes of the correlations. 

Up to now, for simplicity, we assumed that the flow components of hadron momenta are independently smeared only by the hadronisation. There can also be other processes smearing the momenta, including pre-hadronisation random walk and 
post-hadronisation rescattering. The procedure for correcting for smearing presented below can effectively account for all these effects.

Adding to the procedure the transverse momentum magnitudes, \( p_{\textrm{T}, 1} \), and \( p_{\textrm{T}, 2} \), should improve the resolution of the measurement of the flow opening angle. The simulation results, which are not shown here, indicate that the improvement is insignificant. Thus, for simplicity, it is not discussed here in detail.  

\vspace{0.5cm}
\noindent
{\normalfont\bfseries Correcting for smearing. }
\noindent
The hadronisation and other processes significantly smear the distribution of flow momentum components. The latter is given by the wanted spatial correlations of charm and anticharm quarks at the hadronisation (see Appendix~B). Thus, it is important to discuss the possibility of extracting the distribution of $\mathit{\Phi}$ for $c$ and $\bar{c}$ quarks due to flow from the measured distribution for $D$ and $\bar{D}$. 
Let us consider the azimuthal angle of a \( D \) meson as a sum 
\begin{equation}
    \phi = f + h~,
\end{equation}
where $f$ is the angle given by the quark flow and $h$ is its bias due to smearing. From the system isotropy follows that the smearing ($h$) of $c$ and $\bar{c}$ quarks is
independent of the flow azimuthal angle ($f$). Thus, one has,
\begin{equation}
\rho( f_1, f_2, h_1, h_2) = 
    F( f_1, f_2 ) \cdot H( h_1, h_2 )~,
\label{eq:fact}
\end{equation}
where $f_1$, $f_2$ and $h_1$, $h_2$ are flow and hadronisation components of 
the $D$ and $\bar{D}$ azimuthal angle. The functions $\rho(f_1, f_2, h_1, h_2$), $F(f_1, f_2)$ and $H(h_1, h_2)$ are probability density functions of the corresponding random variables.
The smearing by the hadronisation is independent for $c$ and $\bar{c}$ quarks in their respective rest system. However, the Lorentz transformation to the collision centre of mass system using the flow velocity correlates the smearing components in this system, $h_1$ and $h_2$, as, in general, the flow velocities are correlated. Their probability density function results from averaging over the flow velocities:
\begin{equation}
    H( h_1, h_2 ) = \int dv_1\,dv_2\, V( v_1, v_2)\, G( h_1, h_2 | v_1, v_2 )~,
\end{equation}
where the functions $V(v_1, v_2)$ and $G( h_1, h_2 | v_1, v_2)$ are probability densities of the flow velocities and the smearing azimuthal angle components, given the flow velocities.
It is important to note that both $F(..)$ and $H(..)$ distribution functions depend on the space-time correlation model, but on its different features. The $F(..)$ function depends on the wanted azimuthal correlations, and the $H(..)$ function on the radial correlations. The latter introduces a model-dependent smearing, which will be corrected for, see below. Given $F(..)$ and $H(..)$, one can calculate the joint distribution of the azimuthal angles \( J( \phi_1, \phi_2 )\) as:
\begin{equation}
    J(\phi_1,\phi_2)=\int \int \int \int df_1\,df_2\,dh_1\,dh_2\,F(f_1,f_2)\,H(h_1,h_2)\,\delta(f_1+h_1-\phi_1)\,\delta(f_2+h_2-\phi_2)~.
\end{equation}
Then the distribution of the opening angle \( P(\it\Phi) \) can be obtained using Eq.~\ref{eq:Phi}.

The azimuthal angle symmetry implies the symmetry of the marginal functions:
\( H(h_1) = H(-h_1)  \), \( H(h_2) = H(-h_2)  \) and
\( F(f_1) = F(-f_1)  \), \( F(f_2) = F(-f_2)  \).  
Moreover, \(  F(f_1) \) and \(  F(f_2) \) are the uniform distributions between
\(  -\pi \) and \( \pi \).
The factorisation of \( \rho(....) \) (Eq.~\ref{eq:fact}) and the constraints resulting from the symmetry
significantly simplify the procedure of unfolding the wanted distribution
\( F(f_1, f_2\) ) from the measured distribution \( J( \phi_1, \phi_2 )\). 
Still, this requires knowing the smearing function $H(..)$.
There are two obvious options. The smearing function can be calculated from the model. An example would be the statistical hadronisation model as given by Eq.~\ref{eq:had}. The other possibility is to postulate an analytical form of the expected bell-shaped marginal distribution \( H(h) \) with free parameters controlling its form and the correlation between $h_1$ and $h_2$, and use the regularisation methods to extract \( F(f_1, f_2) \) and the parameters of $H(..)$.

\begin{figure}[ht]
\includegraphics[width=0.9\textwidth]{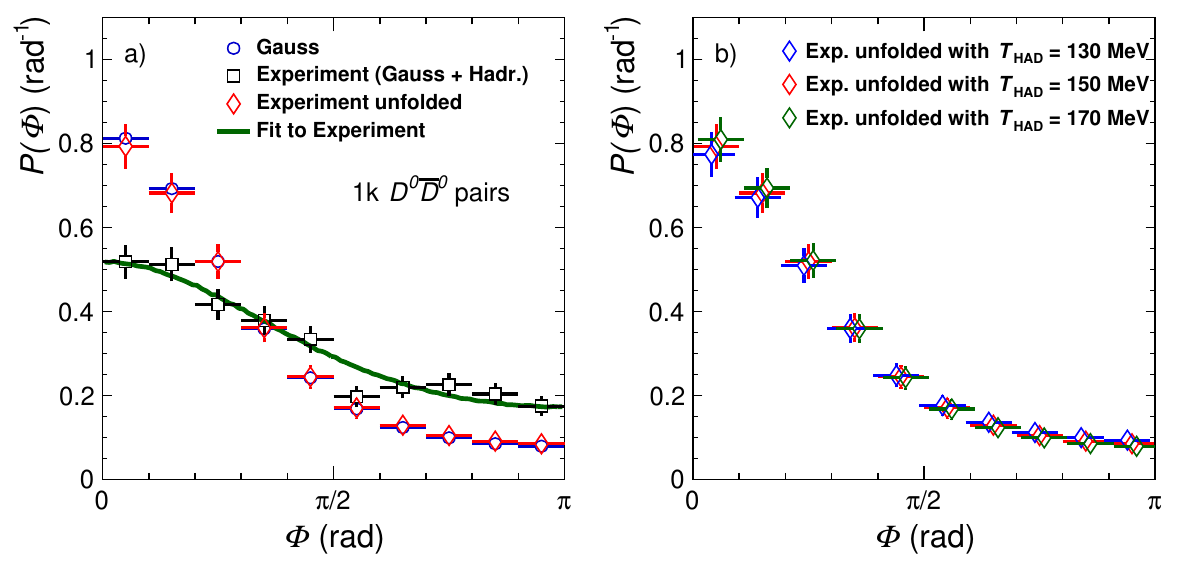}
\caption{
\
The unfolding procedure correcting for smearing of the azimuthal opening angle of $D^0$$\bar{D}^0$ pairs is tested. The left plot shows the sensitivity to the assumed 
value of the spatial correlation parameter, and the right plot shows the sensitivity to the hadronisation parameter. The points are shifted horizontally for clarity. See text for the full description.
}
\label{fig:unfolding:res}
\end{figure}

In the following, we report on testing the unfolding procedure, reducing the general formulation to the 1D unfolding in the azimuthal opening angle \( \it\Phi \). Using the iterative Richardson-Lucy unfolding, we implemented the procedure within the RooUnfold framework~\cite{Brenner:2019lmf}.
For testing, the experimental-like data were generated using the 3D Gauss spatial correlation model with hadronisation for 1000 $D^0$$\bar{D}^0$ pairs, see point $(c)$ above.
The spatial correlation parameter was set to $\sigma = $~2~fm and the hadronisation parameter $T_{\mathrm{HAD}} = $~150~MeV. The corresponding distribution \( P(\mathit\Phi) \) is shown in Fig.~\ref{fig:unfolding:res} (Experiment (Gauss + Hadronisation)) together with the wanted distribution of the flow opening angle (Gauss). 
These distributions correspond to the distributions shown in Fig.~\ref{fig:DPhi}, but Fig.~\ref{fig:unfolding:res} also reflects the finite statistics of the Experiment distribution.

To unfold the smearing due to the hadronisation, the response matrix  
$R(\mathit\Phi_F, \mathit\Phi_E)$ was calculated, where $\mathit\Phi_F$ and $\mathit\Phi_E$ are the opening angles due to the flow only and the flow with hadronisation components added, respectively. We recall that within the 3D Gauss model used for testing, the difference between the angles and thus the response matrix depends on both parameters $\sigma$ and $T_\mathrm{HAD}$. Several tests were performed to check the sensitivity of the unfolded distribution on different settings of $\sigma$ and $T_\mathrm{HAD}$ used to obtain $H(h_1,h_2)$. The closure test done by setting $\sigma = $~2~fm and $T_\mathrm{HAD} = $~150~MeV was passed successfully.
The sensitivity to the assumed $\sigma$ parameter was checked by setting
$\sigma = $~3~fm. It is the upper limit of $\sigma$ calculated assuming absence of the smearing ($H(h_1, h_2) = \delta(h_1) \delta(h_2)$) and requesting that the resulting distribution is close to the generated Experiment distribution (Fit to Experiment in Fig.~\ref{fig:unfolding:res}~(a)). The unfolded distribution is shown in Fig.~\ref{fig:unfolding:res}~(a) (Experiment unfolded), and it matches well the wanted one (Gauss). Finally, we checked the unfolding sensitivity to the $T_\mathrm{HAD}$ parameter by repeating the tests for $T_\mathrm{HAD}$ set to 130 and 170~MeV while fixing $\sigma$ to 3~fm. The unfolded distributions are shown in Fig.~\ref{fig:unfolding:res}~(b) and agree with each other within statistical uncertainties.

\clearpage

\vspace{0.5cm}
\noindent
{\normalfont\bfseries  Collisions with many \( c-\bar{c} \) pairs}
\\
The case with a single $c$ and $\bar{c}$ quark pair produced in individual collisions is the simplest, allowing the study of two-particle spatial correlations unaffected by multi-particle correlations between charm and anticharm quarks or hadrons. This paper focuses on this case. For completeness, we briefly address the case with many \( c-\bar{c} \) pairs produced in individual Pb+Pb collisions.

Let us denote the number of \( c \) quarks in a central Pb+Pb collision by \( N \). Due to charm conservation, the number of \( \bar{c} \) quarks is also \( N \). For simplicity, we assume that all \( c \) and \( \bar{c} \) quarks are measured, and \( N \) is the same for all collisions.

For \( N = 1 \) (the mean \( N \) is close to one at the top CERN SPS energy), the inclusive charm production, averaged over collision properties, is described by the two-particle probability density 
\( J_2(\mathbf{p}_1; \mathbf{p}_2) \). 
For \( N = 30 \) (the mean \( N \) at the CERN LHC), the inclusive charm production is given by the 60-particle density:
\begin{equation}
J_{60}(\mathbf{p}_1, \mathbf{p}_2, \ldots; \ldots, \mathbf{p}_{N+29}, \mathbf{p}_{N+30})~,
\end{equation}
where \( \mathbf{p}_i \) and \( \mathbf{p}_{N+i} \) represent the momenta of a \( c \) and \( \bar{c} \) particles, respectively.

No fundamental laws of physics directly relate \( J_2 \) and \( J_{60} \). Their relationship can only be established by introducing models. A simple model assumes that charm-anticharm pairs are created independently and that particles in a pair are labelled (e.g., each pair is created at a single space-time point, which is measured for each pair). This model results in the factorisation of \( J_{60} \):
\begin{equation}
J_{60}(.;.) = 
J_2(\mathbf{p}_1; \mathbf{p}_{N+1}) 
\cdot J_2(\mathbf{p}_2; \mathbf{p}_{N+2}) 
\cdots J_2(\mathbf{p}_{30}; \mathbf{p}_{N+30})~,
\label{eq:factorization}
\end{equation}
where \( \mathbf{p}_i \) and \( \mathbf{p}_{N+i} \) are the momenta of \( c \) and \( \bar{c} \) particles originating from the same pair. Here, \( J_2(...) \) can be trivially obtained by measuring \( J_2(\mathbf{p}_1; \mathbf{p}_2) \).

Even if the pair labels are unknown, the assumption of independent pair creation enables the extraction of \( J_2(.;.) \) using the balance function method~\cite{Bass:2000az}. This method relies on the distributions of all possible particle pairs with opposite sign (OS) and same sign (SS). 
Here, OS corresponds to \( c-\bar{c} \) and \( \bar{c}-c \) pairs, while SS corresponds to \( \bar{c}-\bar{c} \) and \( c-c \) pairs. The properly normalised differences, OS - SS, can then be identified as \( J_2(...) \), based on the classical-physics-rooted factorisation assumption, Eq.~\ref{eq:factorization}.

However, this factorisation assumption represents a major loophole in interpreting experimental results. The only way to eliminate this loophole is to measure \( J_2(.;.) \) directly, which can be done for collisions with a single $c-\bar{c}$ pair. In practice, they will be approximated by collisions with a negligible contribution from events involving two or more pairs.

%%%%%%%%%%%%%%%%%%%%%%%%%%%%%%%
\section{Feasibility of experimental measurements}
\label{sec:experiment}

This section briefly discusses the requirements for the experimental measurements of correlations between charm and anticharm hadrons produced in central heavy-ion collisions. The important physics condition is a mean multiplicity of charm being small enough to neglect the production of two or more pairs of charm-anticharm hadrons. This requirement implies the measurements at relatively small collision energies, probably close to the top SPS energy of $\sqrt{s_{NN}} \approx 20$~GeV; see Appendix~C. It also suggests collecting data in the fixed target mode, which allows for high detection acceptance and efficiency due to the Lorentz boost of the centre-of-mass. 
For now, we only consider measurements of the most abundant charm and anticharm hadrons, $D^0$ and $\bar{D^0}$ mesons. 
The required statistics of recorded central Pb+Pb collisions can be derived from the average number of reconstructed $D^0\bar{D^0}$-pairs, $\langle D^0 \bar{D^0} \rangle_{\textnormal{rec}}$. In Appendix~C, we estimate that modern experiments at the CERN SPS should be able to record sufficient data to measure 
1000 or more $D^0\bar{D^0}$-pairs.

Figure~\ref{fig:unfolding:res} (a) demonstrates the statistical precision of a signal from 1000 $D^0\bar{D^0}$ pairs, assuming that the statistical fluctuations of background pairs can be neglected. We conclude that this is sufficient to distinguish between the Gauss-like and uniform spatial correlation functions. Moreover, applying the unfolding method (see Sec.~\ref{sec:model}) allows one to remove the smearing due to the hadronisation process, providing access to the spatial correlations between charm and anticharm quarks at hadronisation.

\section{Summary}
\label{sec:summary}
In this work, we propose to study the spatial correlation of charm-anticharm quarks at hadronisation by measuring the momentum correlation of charm-anticharm hadrons produced in central heavy-ion collisions at collision energies with the mean multiplicity of $c-\bar{c}$ pairs below one.

We show that, in particular, the azimuthal correlations of charm and anticharm hadrons observed in an experiment are sensitive to the form of the spatial correlation function of the quarks at hadronisation. 
Furthermore, we discuss the possibility of correcting the results for smearing by the hadronisation and other processes that smear independently the charm and anticharm hadron momenta. 
As a result, the correlation function in the opening angle of flow velocities 
of $c$ and $\bar{c}$ quarks at hadronisation can be obtained.
In the case of central heavy-ion collisions, the correlation function is isotropic in the transverse plane, and it is equal to the spatial correlation function in the azimuthal opening angle between vectors given by the quark hadronisation points.

The experimental results on the spatial correlation function will allow testing different assumptions on the $c$ and $\bar{c}$ quark creation and subsequent transport in the dense medium.
In particular, we hope to explore the possibility of observing the
apparent teleportation of charm and anticharm quarks in the future
publication.

Since the production of multiple pairs of charm and anticharm hadrons in a single collision may spoil the wanted two-particle correlations, it is recommended that the measurements be performed at sufficiently low collision energies, granting a low production probability of multiple charm pairs. 
The proposed method can also be used for hadrons carrying bottom and anti-bottom quarks.

As a quantitative example, we consider charm and anti-charm hadron measurements in central Pb+Pb collisions at the CERN SPS. Assuming typical values of data-taking parameters for the NA61/SHINE experiment at SPS, we show that the required measurements would need
a data-taking rate of 10 kHz or more. These rates are easily allowed by the current detector technologies~\cite{Liu:2024hcq}. Thus, the corresponding measurements may be possible by the upgraded \NASixtyOne~\cite{NA61:2014lfx} and the recently 
proposed DICE/NA60+~\cite{Ahdida:2845241} experiments after the CERN LS3 upgrade period.

\begin{acknowledgments} 
We thank F.\ Giacosa, M.\ Gorenstein and St.\ Mrowczynski for their comments.
This work is partially supported by
the Polish National Science Centre grants 2018/30/A/ST2/00226, 2018/30/E/ST2/00089 and 2020/39/D/ST2/02054.

\end{acknowledgments}
\clearpage

\section*{Data Availability Statement}

This manuscript has no associated data. 

\newpage
\textbf{\Large Appendices}

\vspace{0.2cm}
\textbf{A. Classical vs quantum-mechanical approach to heavy-quark production.} 
Here, we discuss why the classical approximation for charm production in the limit of a single pair is incorrect. The simplest model to address it is hydrodynamics. The main equation is given by the local energy-momentum conservation, which in terms of the expectation values of the energy-momentum tensor, $T^{\mu\nu}$, reads:
\begin{equation}
    0= \partial_\mu T^{\mu\nu} = \partial_\mu {\rm tr}\left( \hat \rho \, \widehat T^{\mu\nu} \right)~,
\end{equation}
with respect to the density matrix $\hat \rho$ of the system. Additional equations, the equation of state and the treatment of the non-ideal part (transport coefficients) allow us to solve the system for the expectation values $T^{\mu\nu}$. Statistical hadronisation is then used to calculate predictions for particle production. Scattering after the hadronisation is usually considered with a separate transport phase. In some cases, the baryon number conservation equation
\begin{equation}
    0=\partial_\mu J^\mu_B = \partial_{\mu}{\rm tr} \left( \hat \rho \, \widehat J^\mu_B \right),
\end{equation}
is added to the hydrodynamics equations. In principle, the electric current and the other conserved charges should also be considered when calculating the charge densities. Moreover, the link between tensors in space-time and particles in phase space, necessary at the hadronisation stage, is through the relativistic Wigner distribution $W(x,p)$~\cite{DeGroot:1980dk}. The latter is the generalization of the classical distribution function $W(x,p)\xrightarrow{\rm classical \; limit}\propto \delta(p^2-m^2)f(x,{\bf p})$. It depends on the bi-linearity of the fields and the one-particle reduced density matrix. It does not depend on the two-particle ones and higher orders. By construction, regardless of the ansatz (local equilibrium, viscous corrections), the hadronisation formula is for the one-particle observables only. All of the content about particle correlations must come from somewhere else.

Despite being a significantly different model, similar considerations hold for the relativistic Boltzmann equation because it also stems from $W(x, p)$. Relativistic kinetic theory is a limit of the evolution of the Wigner distribution. As explained in Ref.~\cite{DeGroot:1980dk}, the approximations needed to use the relativistic Boltzmann equation instead of the more general equations for the evolution of $W(x,p)$ include both arbitrarily small gradients and arbitrarily weak interactions. Then, one can neglect the coupling with the two-particle reduced density matrix, and the only quantum leftover is the cross-section, which must be evaluated in the framework of axiomatic field theory. These two conditions are enough to question whether the relativistic kinetic theory can be used for the QGP. Strong interactions and large gradients are needed to fit the experimental data on top of a realistic (non-ideal) state equation that already requires phenomenological modifications to the simple relativistic Boltzmann equation. All the phenomenological modifications used in the state-of-the-art models (temperature-dependent masses, off-shell cross sections, etc.) do not insert any contribution from the $n$-particle reduced density matrix. This sector of the microscopic theory is systematically neglected. The spectra can be deduced from $W(x,p)$ alone, a one-particle object. If the evolution of $W(x,p)$ couples mostly to itself, one can argue that these extensions of the relativistic kinetic theory have a good chance to reproduce the spectra (and $v_2$ and other one-particle objects). Still, the same cannot be said about correlations.

If one prefers a more intuitive approach to quantum fields, some considerations must be made from first principles. Because of the Heisenberg uncertainty principle, one cannot have an arbitrarily sharp wave function in both position and momentum at the same time. The more the $c -\bar{c}$ pair is well-defined in momentum, the more it must be delocalized. Suppose the quarks are assumed to be produced as close as possible to momentum eigenstates, to forget about the details of the wave function in momentum space. In that case, one has to consider them substantially delocalized in space. They cannot be considered in a single cell, and the wave function in the configuration space gives a weight regarding which part of the medium impacts more the heavy quarks. In any case, neither hydrodynamics nor kinetic theory can treat such wave functions dynamically.

\begin{figure} 
    \includegraphics[width=0.8\textwidth]{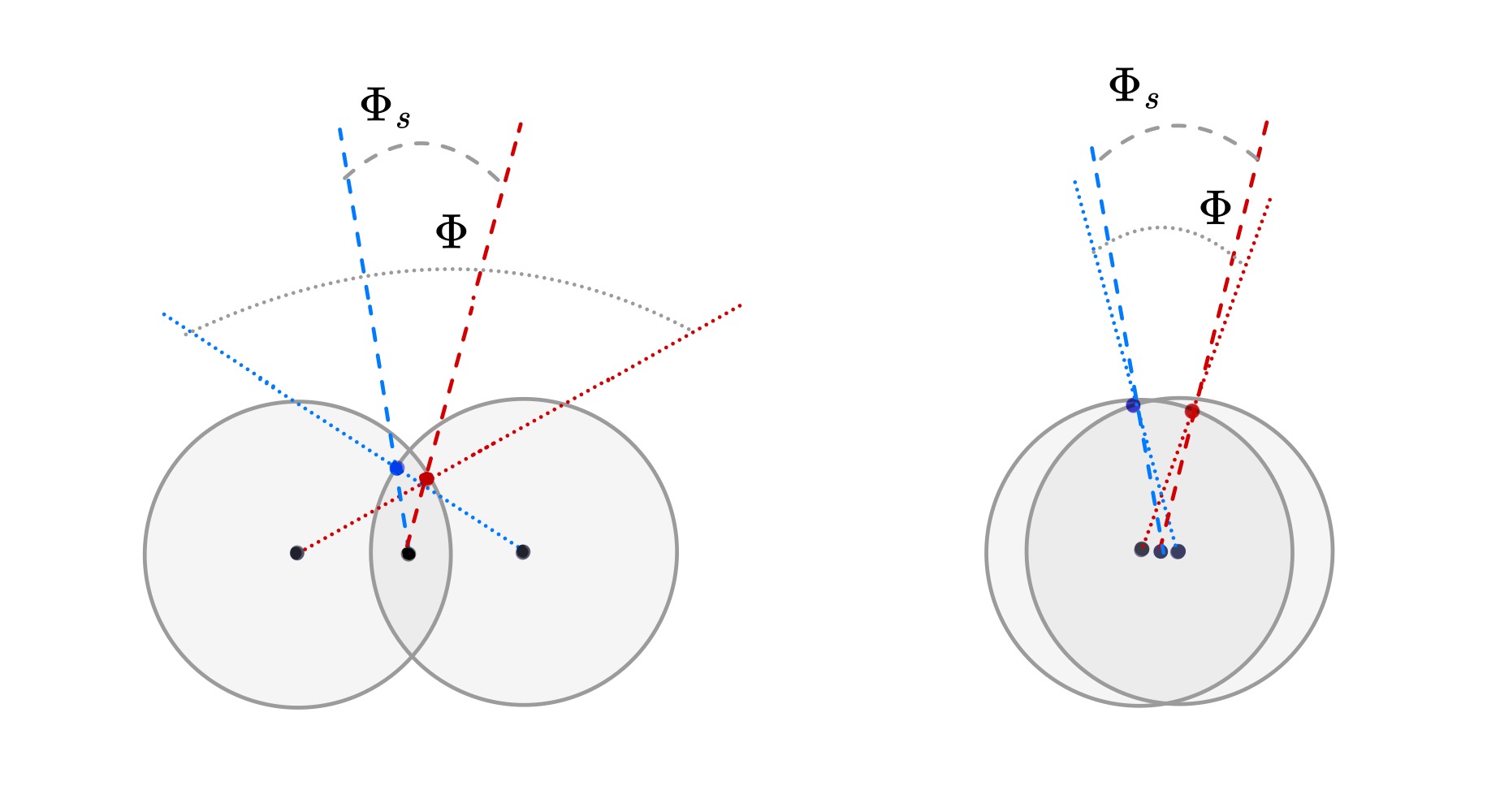}
    \caption{Sketches of two examples of heavy-ion collisions in the plane transverse to the beam direction. The grey circles indicate the two colliding nuclei, and the left and right black dots show their centres. The overlap of two circles gives the initial shape of the collision zone, and the middle black dot indicates its centre. The flow is perpendicular to the surface of the collision zone (dotted lines). The flow approximately preserves the initial shape of the collision zone. Thus, to simplify the plot, the shape of the hadronisation zone is not plotted, and instead, the $c$ and $\bar{c}$ hadronisation points are projected to the initial surface of the collision zone (red and blue dots). The measurable  flow opening angle, the angle between dotted blue and red lines, is denoted as
    $\mathit{\Phi}$, whereas the wanted spatial opening angle, the angle between dashed blue and red lines, is indicated as $\mathit{\Phi}_S$. 
    The geometry sketch for collisions with a large distance between the circle's centres (peripheral collisions) is shown on the left, and for collisions with a small distance (central collisions) on the right.
    Obviously, for the central collisions in the limit of the full overlap, the spatial opening angle is equal to the flow opening angle independently of the orientation of the hadronisation vectors. The relation between $\mathit{\Phi}_S$ and $\mathit{\Phi}$ for peripheral collisions depends on the flow angle and the orientation of the hadronisation vectors relative to the impact parameter. See text for further discussion.
    }
\label{fig:central}
\end{figure}

\vspace{0.2cm}
\textbf{B. Importance of central heavy-ion collisions.}
Here, we discuss the importance of studying spatial correlations of charm and anticharm quarks at hadronisation for central heavy-ion collisions being isotropic in the transverse plane.
In this case, the wanted spatial opening azimuthal angle equals the measured flow opening angle. This is illustrated in Fig.~\ref{fig:central}, where 
the collision geometry in the transverse plane relative to the beam direction is sketched for two examples. 
The left sketch shows the collision of equal nuclei with a large distance between their centres, the so-called peripheral collision.
The right plot depicts the collision with the small distance between the centres,
the so-called central collision. It is easy to see that in the limiting case of zero distance, the central collision, the spatial opening angle $\mathit{\Phi}_S$ is equal to the measured flow opening angle, $\mathit{\Phi}$, of the $c-\bar{c}$ pair at hadronisation. This is not true for peripheral collisions.
In the latter case, the spatial opening angle depends on the flow opening angle and the orientation of the pair relative to the vector connecting the nuclear centres - the impact parameter. Thus, extending and applying the method to peripheral collisions would make the results model-dependent and require high statistics for the measurements relative to the impact parameter.

\vspace{0.2cm}
\textbf{C. Example estimate of event statistics and data-taking time.}
Here, we present a simple estimate of the event statistics and data-taking time assuming detector setup and performance similar to the \NASixtyOne\ experiment at CERN~\cite{NA61:2014lfx} recording Pb+Pb collisions at $\sqrt{s_{NN}} = 17.3$~GeV. Assuming that processes that impact the reconstruction of $D^0$ and $\bar{D^0}$ mesons are approximately uncorrelated, we estimate the average number of reconstructed pairs as
\begin{equation} \label{eq:avgD0D0yield}
    \langle D^0\bar{D^0} \rangle_{rec} \approx
        \langle c\bar{c} \rangle \cdot
        \left(
            P(c \rightarrow D^0) \cdot
            \textnormal{BR}(D^0 \rightarrow K \pi) \cdot
            P(\textnormal{acc}) \cdot
            P(\textnormal{sel}) \cdot
            P(\textnormal{rec})
        \right)^2,
\end{equation}
where $\langle c\bar{c} \rangle$ is the average number of $c-\bar{c}$-pairs per event. The $P(c \rightarrow D^0)$ =  0.31 is a probability for $c$-quark to hadronize into the $D^0$ meson evaluated within the PHSD model \cite{Cassing_2009}, BR$(D^0 \rightarrow K^+ \pi^-)$ =  3.98\% is a branching ratio of decay channel used in the measurements~\cite{Workman:2022ynf}, $P(\textnormal{acc})$ = 0.5 is a probability for $D^0$ to be within an acceptance region of the detector, $P(\textnormal{sel})$ = 0.2 is a probability for $D^0$ to pass background-suppressing selection of charm meson candidates, and $P(\textnormal{rec})$ = 0.9 is a probability of reconstructing the meson.
The value of $P(\textnormal{acc})$ was evaluated using the \GeantFour simulation with the detector setup for November 2022, $P(\textnormal{sel})$ is taken from the pilot analysis of $D^0$ and $\bar{D^0}$ production~\cite{Merzlaya:2771816}, and $P(\textnormal{rec})$ was obtained from a \GeantFour simulation with the setup for November 2022 and reconstruction software used for previous open charm analysis using 2017 and 2018 data ~\cite{mbajda, Merzlaya:2771816}.

Finally, given $\langle D^0\bar{D^0} \rangle_{rec}$, an estimate of the required event statistics can be obtained as
\begin{equation} \label{eq:nevents}
    \{\textnormal{number of central events to collect}\} \approx
        \frac{\{\textnormal{number of $D^0\bar{D^0}$ pairs to reconstruct}\}}
             {\langle D^0\bar{D^0} \rangle_{rec}}.
\end{equation}

The $\langle c\bar{c} \rangle$ value is neither reliably predicted by models nor measured by experiments. However, considering available estimates~\cite{Snoch:2018nnj}, we expect that the value of $\langle c\bar{c} \rangle$ for central Pb+Pb at $\sqrt{s_{NN}} \approx 17$~GeV should range from  0.1 up to 1. 

Putting all together, the run time needed to collect 1000 $D^0\bar{D^0}$-pairs
by the upgraded \NASixtyOne detector is given in Table~\ref{tab:times}. 
The time is calculated assuming different event rates (number of recorded events per second during the spill) and mean multiplicities of charm and anticharm pairs, $\langle c\bar{c} \rangle$.

\begin{table}[h]
\begin{tabular}{|c|c|c|c|c|}
    \hline
              & $\langle c\bar{c} \rangle = 0.1$         & $\langle c\bar{c} \rangle = 0.2$          & $\langle c\bar{c} \rangle = 0.5$     & $\langle c\bar{c} \rangle = 1$           \\ \hline
    1 kHz     & $300$ days & $150$ days   & $62$ days  & $30$ days  \\ \hline
    10 kHz    & $30$ days  & $15$ days    & $6$ days   & $3$ days   \\ \hline
    100 kHz   & $3$ days   & $1$ day     & $<1$ day    & $<1$ day     \\ \hline
            \hline
    $N_{pair}/N_{comb}$ & 91\% & 83\%  & 66\%             & 50\%             \\ \hline
\end{tabular}
\vspace*{0.3cm}
\caption{\label{tab:times} Estimate of the duration of a data-taking period needed to collect 1000 $D^0\bar{D^0}$-pairs (first three rows). 
The event rate is in spills, and in the run time calculations, the accelerator duty cycle of 30\% is assumed.
The last row shows the ratio of the produced pairs of $c-\bar{c}$ quarks to all combinations of them, assuming an independent pair production model.  See text for details. 
}
\end{table} 

A typical ion beam period at CERN is about four weeks.
Entries in Table~\ref{tab:times} with a data-taking time of 100 days or more correspond to scenarios where the measurement may take longer than a period between
the CERN accelerators' long shutdowns. Moreover, at the moment, the event rate of 100~kHz would require a significant upgrade of the \NASixtyOne detector and its beamline. However, a setup corresponding to 10~kHz may be achievable within the next years. 
We note that the ALICE experiment at the CERN LHC records Pb+Pb collisions with a rate of about 50~kHz~\cite{Liu:2024hcq} using the same type of silicon pixel detectors as the ones installed in \NASixtyOne.
An additional possibility for the experimental study would be constructing a new experiment optimised for charm measurements. The DICE/NA60+ proto-collaboration submitted recently a letter of intent to the CERN SPSC~\cite{Ahdida:2845241}. 

Thus, we find that having $\langle c\bar{c} \rangle > 0.2$ in central Pb+Pb collisions at the top CERN SPS energy, it should be possible to perform the measurements of $c-\bar{c}$ correlations at the CERN SPS rather soon.

\bigskip\newpage
\bibliographystyle{utphys}
\bibliography{references}

\end{document}